\documentclass[
11pt,tightenlines,%
nofootinbib,preprintnumbers,prd]{revtex4}
\usepackage[T1]{fontenc}
\usepackage[latin1]{inputenc}
\usepackage{graphics}
\usepackage{eepic}
\usepackage{epsfig}
\usepackage{amssymb}

\begin{document}
\preprint{TUW-03-11}
\title{HTL 
quasiparticle models 
of deconfined QCD\\
at finite chemical potential}
\author{A. Rebhan}
\affiliation{Institut f\"ur Theoretische Physik, Technische
Universit\"at Wien, 
A-1040 Vienna, Austria }
\author{P. Romatschke}
\affiliation{Institut f\"ur Theoretische Physik, Technische
Universit\"at Wien, 
A-1040 Vienna, Austria }

\begin{abstract}
Using quasiparticle models and imposing
thermodynamic consistency, lattice data for the
equation of state of deconfined QCD
can be mapped to finite chemical potential. 
We consider a refinement of existing simple massive quasiparticle models
using the non-local hard-thermal-loop (HTL) propagators, and
certain NLO corrections thereof, to obtain the thermodynamic
potential as a function of temperature and chemical potential.
At small chemical potential we find
that the results for the slope of constant pressure from our main model 
for 2 massless quark flavors is
in good agreement with recent lattice data for 2+1 flavors while it
deviates somewhat from 
lattice data for 2 flavors from another group.
For zero temperature, we obtain an estimate
for the critical chemical potential which is
close to that obtained from simpler quasiparticle models.
\end{abstract}
\maketitle

\section{Introduction}

Asymptotic freedom suggests that at sufficiently high temperature
and/or quark chemical potential QCD is deconfined, i.e. can be
described in terms of the fundamental quark and gluonic degrees
of freedom \cite{Collins:1975ky}. 
High temperature and small chemical potential are of relevance
to the quest for the quark-gluon plasma in heavy-ion experiments;
sufficiently high chemical potential (at
comparatively low temperature) may be reached in the cores of
compact stars. In the latter case, novel color superconducting
phases may occur \cite{Alford:2002wf}, 
which should however have only 
comparatively small effects on the equation of state.

The thermodynamic pressure of QCD at high temperature $T\ll \Lambda_{\rm QCD}$ 
has been determined in perturbation theory to order
$\alpha^{5/2}_{s}$ in QCD \cite{
Arnold:1995eb,Zhai:1995ac} and 
even to 
$\alpha^{3}\log{\alpha}$ in a recent heroic effort \cite{Kajantie:2002wa}.
Strict perturbation theory, however, shows extremely poor convergence
for any temperature of practical interest so that further resummations
appear to be necessary. 
In recent years, resummations which are based on
the hard-thermal-loop (HTL) effective action 
\cite{Braaten:1992gm
} have been proposed,
alternatively in form of so-called HTL perturbation theory
\cite{Andersen:1999fw,
Andersen:2002ey
}
or based upon the 2-loop $\Phi$-derivable approximation 
\cite{Blaizot:1999ip
,Blaizot:2000fc,Blaizot:2003tw}
(see also Peshier \cite{Peshier:2000hx}).
The latter approach, which assumes weakly interacting
quasiparticles as determined by the HTL propagators
and NLO corrections thereof, leads to results which
agree remarkably well with lattice data down to about $3 T_c$
if standard 2-loop running $\alpha_s$ is adopted,\footnote{%
This is in fact not the case for 2-loop HTL perturbation theory
\cite{Andersen:2002ey
}, but it appears that
the source of the difficulty has to do with the necessity of
thermal counter-terms and incompletely compensated hard contributions,
see Ref.~\cite{Blaizot:2003iq}.}
and it has been recently shown that these results are in fact
consistent with those obtained in high-order perturbation theory if the
latter is organized through effective (dimensionally reduced)
field theory according to \cite{Braaten:1996jr} and effective-field-theory
parameters are kept without further expanding in powers of the
coupling \cite{Blaizot:2003iq}, as already 
advocated in Ref.~\cite{Kajantie:2002wa}.

The behavior of the thermodynamic potentials closer to the
transition temperature needs to be
studied by fully non-perturbative means.
For zero quark chemical potential $\mu=0$ lattice QCD calculations are 
seemingly up to the task of determining thermodynamic quantities
 for a quark-gluon plasma very accurately by now 
\cite{Karsch:2001cy,Laermann:2003cv}.
Although recently 
there has been important progress also for non-vanishing $\mu$ 
\cite{Fodor:2001au,deForcrand:2002ci,Fodor:2002km,Allton:2002zi,D'Elia:2002gd,Gavai:2003mf},
the case of large $\mu \gg T$ 
(relevant for the equation of state for cold dense matter, 
which is of importance in astrophysical 
situations \cite{Peshier:1999ww,Peshier:2002ww,Fraga:2001xc,%
Andersen:2002jz})
is currently beyond the reach of lattice calculations.

As a remedy for this situation, Peshier 
\emph{et al.} \cite{Peshier:1999ww,Peshier:2002ww} proposed a method which
can be used to map the available lattice data for $\mu=0$ to finite $\mu$
and small temperatures by describing the interacting plasma as a system
of massive quasiparticles (QPs); section 2 gives a short review of
this technique. While in their simple QP model
the (thermal) masses of the quarks and gluons are approximated
by the asymptotic limit of the hard-thermal-loop (HTL) self-energies, 
we consider models based on the HTL-resummed 
entropy \cite{Blaizot:1999ip,Blaizot:2000fc,Blaizot:2003tw}, 
which include more of the 
physically important plasmon effect than in the simple QP model,
and (NLO) extensions thereof, which include the full plasmon effect.
In perturbation theory it is essentially this effect that spoils
convergence, so a priori it was not granted that a model including this
effect fully would give physically sensible results. As it turns out, 
however, the results based on our models (which are presented in section
3 and 4, respectively) turn out to be well-behaved and indeed lie not 
too far from the simple QP model.

In section 5 we give the 2-parameter fits of the various models to 
the lattice data \cite{AliKhan:2001ek} for 2 flavors and compare 
the model results for the pressure at small temperatures 
and large chemical potential in section 6. Furthermore, we compare 
our results for the isobar emerging at\footnote{Here and in the following
$T_{c}$ is always to be understood as $\left.T_{c}\right|_{\mu=0}$.}
 $T_{c}$ to recent lattice results
and give our conclusions in section 7.

\section{Mapping lattice data to finite chemical potential using QP models}

Assuming an SU($N$) plasma of gluons and $N_{f}$ light quarks
in thermodynamic equilibrium can be described as a weakly interacting gas of 
massive quasiparticles with residual interaction $B$, the pressure 
of the system is given by \cite{Peshier:1999ww}
\begin{equation}
P(T,\mu)=\sum_{i=g,q} p_{i}(T,\mu_{i},m_{i}^2)-B(m_{g},m_{q}),
\label{Pressureansatz}
\end{equation}
where the sum runs over gluons (g), quarks and anti-quarks (q) with 
respective chemical potential $0,\pm\mu$; $p_{i}$ are
the model dependent QP pressures and $m_{i}$ are the QP masses.
The latter 
are functions of an effective strong coupling $G^2(T,\mu)$, which
appears only within $m_i(G^2,T,\mu)$ in a predefined form.

Using the stationarity of the thermodynamic potential 
under variation of the self-energies and Maxwell's 
relations \cite{Peshier:1999ww}, one obtains
a partial differential equation for $G^2$,
\begin{equation}
a_{T} \frac{\partial G^2}{\partial T} + a_{\mu} \frac{\partial G^2}{\partial %
\mu} = b,
\label{Andrefloweq}
\end{equation}
where $a_{T}$, $a_{\mu}$ and $b$ are coefficients that are given by integrals 
depending on $T,\mu$ and $G^2$.
Given a valid boundary condition, a solution
for $G^2(T,\mu)$ is found by solving the above flow equation by 
the method of characteristics;
once $G^2$ is thus known in the $T,\mu$ plane, the 
QP pressure is fixed completely.
The residual interaction $B$ is then given by the integral
\begin{equation}
B=\int \sum_{i} \frac{\partial p_{i}}{\partial m_{i}^2}\left(\frac{\partial %
m_{i}^2}{\partial \mu} d\mu + \frac{\partial m_{i}^2}{\partial T} dT\right)%
+B_{0},
\label{Bdet}
\end{equation}
where $B_{0}$ is an integration constant that has to be fixed by lattice
data (usually by requiring $P(T_{c},\mu=0)=P_{lattice}(T_{c})$).
Motivated by the fact that at $\mu=0$ and $T\gg T_{c}$ the
coupling should behave as predicted by the perturbative QCD beta-function,
Ref. \cite{Peshier:1999ww} used the ansatz
\begin{equation}
G^2(T,0)=\frac{48 \pi^2}{(11N-2N_{f})\ln{\frac{T+T_{s}}{T_c}\lambda}}.
\label{G2ansatz}
\end{equation}
The parameters $\lambda$ and $T_{s}$ are determined by fitting the entropy 
of the model (which is independent of $B$)
to available lattice data at $\mu=0$. Using (\ref{G2ansatz}) as boundary
condition for (\ref{Andrefloweq}), the above procedure allows one to 
map the lattice pressure from $\mu=0$ to the whole $T,\mu$-plane.

\section{Simple and HTL QP model}

A simple ansatz for the QP pressure is obtained by taking $p_{i}$ to be 
the pressure of a free gas of massive particles
\cite{Peshier:1999ww,Peshier:2002ww}, with the masses defined 
as the asymptotic (i.e. large momentum) gluonic and fermionic self-energies,
respectively \cite{Flechsig:1996ju}:
\begin{eqnarray}
\label{minfty2}
\hat m_\infty^2&=&(2N+N_{f})\frac{G^2 T^2}{12}+N_{f}%
\frac{G^2 \mu^2}{4 \pi^2},\\
\label{Minfty2}
\hat{M}_\infty^2&=&\frac{G^2 (N^2-1)}{8 N}\left(T^2+\frac{\mu^2}{\pi^2}\right).
\end{eqnarray}

Comparing with perturbation theory
one finds that while the Stefan-Boltzmann and leading-order interaction
terms are correctly reproduced, 
only $1/(4\sqrt{2})$ of the NLO term of the interaction
pressure (the plasmon effect $\sim G^3 T^4$) 
is present \cite{Blaizot:2000fc}.

To include more of this physically relevant effect, one can make 
use of the HTL-resummed 
entropy \cite{Blaizot:1999ip,Blaizot:2000fc,Blaizot:2003tw}, which takes into
account the momentum dependence of the QP excitations as well as Landau-damping
effects. The HTL model ansatz for the QP pressure for gluons ($p_{g}$)
and quarks $p_{q}$ then reads
\begin{eqnarray}
p_g & = & - d_{g} \int \frac{d^3 k}{(2\pi)^3} \int _{0}^{\infty}%
 \frac{d\omega}{2\pi} n(\omega)%
\left[2 \rm{Im} \ln{\left(-\omega^2+k^2+\hat{\Pi}_{T}\right)}%
-2\rm{Im} \hat{\Pi}_{T} \rm{Re} \hat{D}_{T} \right. \nonumber \\
& & \left. + \rm{Im} \ln{\left(k^2+\hat{\Pi}_{L}\right)} +%
\rm{Im} \hat{\Pi}_{L} \rm{Re} \hat{D}_{L}\right] \label{pg} \\
p_q & = & - d_{q} \int \frac{d^3 k}{(2\pi)^3} \int_{0}^{\infty} %
\frac{d \omega}{2 \pi} \left(f_{+}(\omega)+f_{-}(\omega)\right)%
\left[\rm{Im}\ln{\left(k-\omega+\hat{\Sigma}_{+}\right)} \right. \nonumber\\
& & \left.-\rm{Im}\hat{\Sigma}_{+}\rm{Re}\hat{\Delta}_{+}%
+ \rm{Im}\ln{\left(k+\omega+\hat{\Sigma}_{-}\right)}+%
\rm{Im}\hat{\Sigma}_{-}%
\rm{Re}\hat{\Delta}_{-}\right],
\label{pq}
\end{eqnarray}
where $d_{g}=2(N^2-1)$, $d_{q}=2 N N_{f}$ for gluons and 
quarks/anti-quarks, respectively; $n(\omega)$ and $f_{\pm}(\omega)$
are the bosonic and fermionic distribution functions.
$\hat{D}_{T,L}$, $\hat{\Delta}_{\pm}$ 
are the HTL propagators with $\hat{\Pi}_{T,L}$ and $\hat{\Sigma}_{\pm}$
the corresponding self-energies,
\begin{eqnarray}
\hat{\Pi}_{L}(\omega,k)& = & \hat{m}%
_{D}^2\left[1-\frac{\omega}{2 k} \ln{\frac{%
\omega+k}{\omega-k}}\right]  \label{PiL}\\
\hat{\Pi}_{T}(\omega,k)& = &\frac{1}{2}\left[\hat{m}_{D}^2+%
\frac{\omega^2-k^2}{k^2}%
\hat{\Pi_{L}}\right] \label{PiT}\\
\hat{\Sigma}_{\pm}& = & \frac{\hat{M}^2}{k}\left[1-\frac{\omega\mp k}{2k}%
\ln{\frac{\omega+k}{\omega-k}}\right]; \label{Sig}
\end{eqnarray}
which we have written in terms of the HTL Debye mass squared
$\hat m_{D}^2=2 \hat m_\infty^2$ and the HTL zero-momentum 
quark mass squared
$\hat{M}^2=\hat M_\infty^2/2$.

The
HTL QP model includes 
$1/4$ of the full plasmon effect at $\mu=0$, 
which is a factor of $\sqrt2$ more than in the simple QP model
which only uses the asymptotic HTL masses. 
The remainder of the plasmon effect is in fact entirely due to NLO
corrections to these asymptotic masses at order
$G^3T^2$ \cite{Blaizot:1999ip,Blaizot:2000fc,Blaizot:2003tw}.

\section{NLO models}

The complete (momentum-dependent)
NLO corrections to the asymptotic masses
of quarks and gluons have not been calculated yet, but
as far as the plasmon effect is concerned, these corrections
contribute in the averaged form \cite{Blaizot:1999ip,Blaizot:2000fc}
\begin{equation}\label{deltamasav}
\bar\delta m_\infty^2={\int dk\,k\,n'(k) {\rm Re}\, \delta\Pi_T(\omega=k)
\over \int dk\,k\,n'(k)}=-{1\over2\pi}G^2NT\hat m_D
\end{equation}
and similarly
\begin{equation}\label{deltaMasav}
\bar\delta M_\infty^2={\int dk\,k\,(f'_+(k)+f'_-(k)) {\rm Re}\,
2k \delta\Sigma_+(\omega=k)
\over \int dk\,k\,(f'_+(k)+f'_-(k))}=-{(N^2-1)/(2N)\over 2\pi}G^2T\hat m_D\;.
\end{equation}

For the values of the coupling $G$ considered here, 
these corrections are so large that they 
give tachyonic masses when treated strictly perturbatively.
In Ref.~\cite{Blaizot:2000fc} it has been proposed to incorporate
these corrections through a quadratic gap equation
which works well as an approximation
in the exactly solvable scalar O($N\to\infty$)-model, where
strict perturbation theory would lead to identical difficulties.
However, for the fermionic asymptotic masses,
in order to have the correct scaling of Casimir factors in
the exactly solvable large-$N_f$ limit of QCD
\cite{Moore:2002md
}, a corresponding gap equation has
to remain linear in the fermionic mass squared. 
Choosing the correction
therein to be determined by the solution to the gluonic gap
leads to \cite{Rebhan:2003fj,Blaizot:2003tw}
\begin{eqnarray}
\label{minftyb2}
\bar{m}_{\infty}^2& =& \hat{m}_\infty^2-\frac{G^2 N T}{\sqrt{2} \pi} %
\bar{m}_{\infty} \\
\label{Minftyb2}
\bar{M}_{\infty}^{2}&=& \hat{M}_\infty^2-\frac{G^2 (N^2-1)T }{2 \sqrt{2} \pi N}\bar{m}_{\infty} ,
\end{eqnarray}
where $\hat{m}_\infty^2$ and $\hat{M}_\infty^2$ are the leading-order
gluonic and fermionic asymptotic
masses as given in (\ref{minfty2}) and (\ref{Minfty2}).
This in fact avoids tachyonic masses for the fermions
as long as $N_f\le 3$.
\footnote{%
For $N_f=3$ the solutions to the above approximate gap equations happen to
coincide with those obtained in
the original version of two independent quadratic
gap equations of Ref.~\cite{Blaizot:2000fc}. For the case $N_f=2$
considered in this paper, the differences are fairly small.
For $N_f>3$, however, the necessity to avoid tachyonic masses would
restrict the range of permissible coupling strength $G$.}

Finally, since the averaged quantities $\bar m_\infty^2$
and $\bar M_\infty^2$ are the effective masses at
hard momenta only, a cutoff scale 
$\Lambda=\sqrt{2 \pi T \hat{m}_{D} c_{\Lambda}}$
is introduced that separates soft from hard momenta. 
The QP pressure for this model, which in the following will be referred to
as NLA-model, then separates into a soft and a hard component for both
gluons and fermions. The soft parts are given by expressions like
(\ref{pg}) and (\ref{pq}), but with $\Lambda$ as upper limit for 
the momentum integration. For the hard parts, the momentum 
integrations run from $\Lambda$ to $\infty$ and the mass pre-factors
in the HTL self-energies (\ref{PiL},\ref{PiT},\ref{Sig})
are replaced by their asymptotic counterparts, 
$\hat{m}_{D}^2\rightarrow 2 \bar m_{\infty}^2$ and $\hat{M}^2\rightarrow %
\frac{1}{2} \bar M_{\infty}^2$. 

The single free parameter $c_{\Lambda}$ in $\Lambda$ can be varied around
$1$ to obtain an idea of the ``theoretical error'' of the model. In the 
following we will consider the range
$c_{\Lambda}=\frac{1}{4}$ to $c_{\Lambda}=4$; note that 
$c_{\Lambda}=\infty$ corresponds to the HTL-model 
(which has been used as cross-check) since all hard corrections are ignored.
On the other hand, $c_{\Lambda}=0$ would assume that 
(\ref{minftyb2}) and (\ref{Minftyb2}) represent good approximations
for the NLO corrections to the spectral properties of soft
excitations. However, the few existing results, in particular on
NLO corrections to the Debye mass \cite{Rebhan:1993az
} and 
the plasma frequency \cite{Schulz:1994gf}, appear to be rather different
so that it seems safer to leave the soft sector unchanged by keeping
a finite $c_\Lambda$.

\section{Entropy and pressure for \protect$\mu=0$}

\begin{figure}
\begin{minipage}[t]{.48\linewidth}
\includegraphics[width=\linewidth]{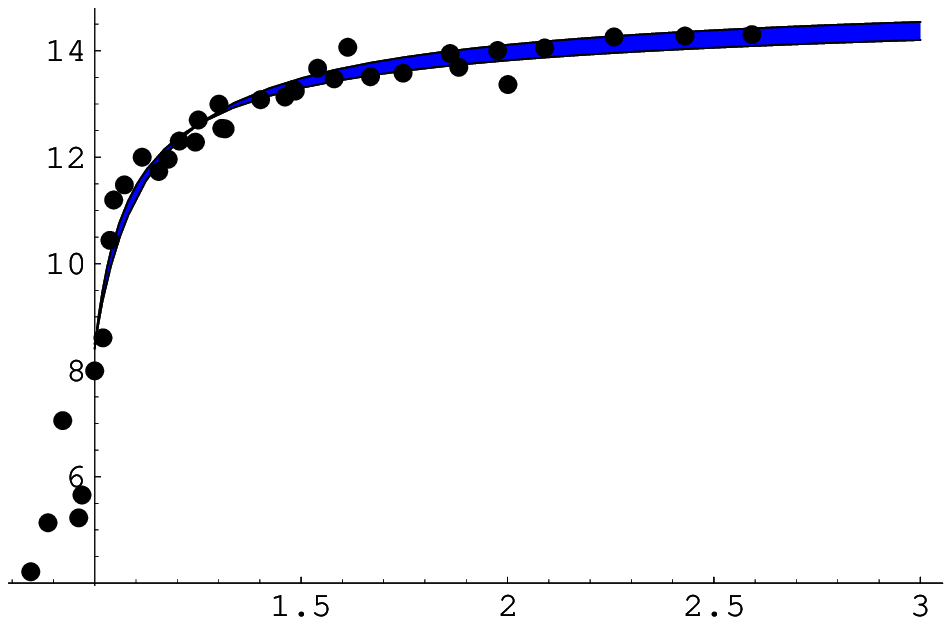}
\setlength{\unitlength}{1cm}
\begin{picture}(6,0)
\put(6.5,0.1){\makebox(0,0){\footnotesize $T/T_{c}$}}
\put(0,5.7){\makebox(0,0){\footnotesize $S/T^3$}}
\end{picture}
\caption{Entropy data generated 
from Ref.~\cite{AliKhan:2001ek} vs.\ fitted model entropy.
The band was generated from NLA model for $c_{\Lambda}$ between 
$4$ and $1$; the HTL and simple QP models lie at the lower boundary.}
\label{fig:S1}
\end{minipage}\hfill
\begin{minipage}[t]{.48\linewidth}
\includegraphics[width=\linewidth]{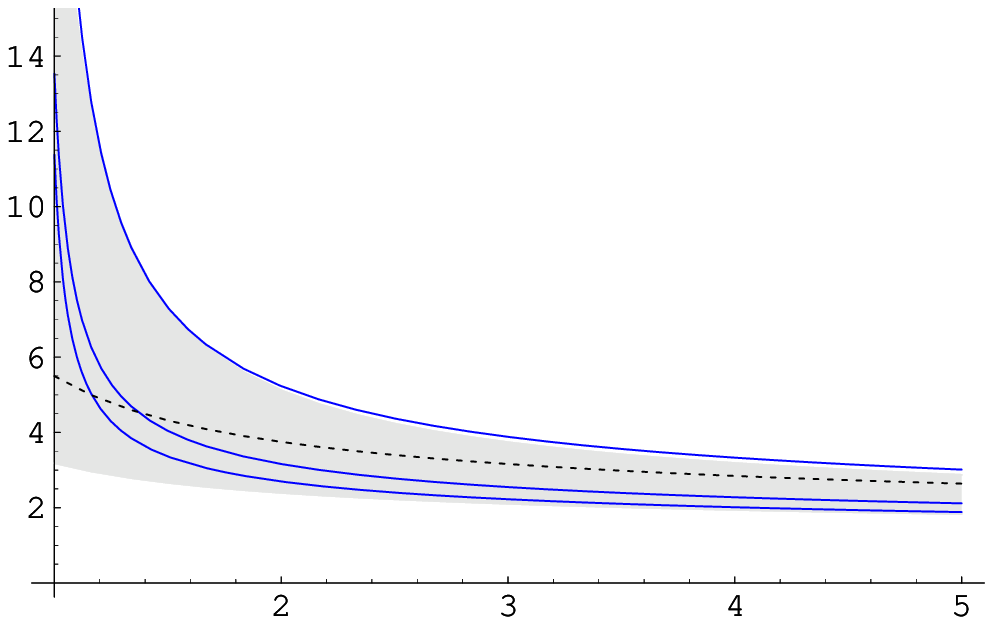}
\setlength{\unitlength}{1cm}
\begin{picture}(5,0)
\put(6.5,0.1){\makebox(0,0){\footnotesize $T/T_{c}$}}
\put(-0.5,5.7){\makebox(0,0){\footnotesize $G^2(T,\mu=0)$}}
\end{picture}
\caption{Effective coupling: NLA model results
for $c_{\Lambda}=$ $4$,$1$ and $1/4$ (full lines from
bottom to top), 2-loop perturbative coupling in 
\protect$\overline{\hbox{MS}}$ (gray band) and result 
from Ref.\protect{\cite{Schneider:2003uz}} (dotted line).}
\label{fig:G1}
\end{minipage}\hfill
\end{figure}

To obtain the input parameters $T_{s}$ and $\lambda$ we fitted the entropy
expressions
 from the models under consideration to lattice data \cite{AliKhan:2001ek}
for $N_{f}=2$, with an estimated continuum extrapolation 
as used in Ref.~\cite{Peshier:2002ww}. 
We find\\
\begin{center}
\begin{tabular}[c]{|c|c|c|c|c|}
\hline
& HTL & $c_{\Lambda}=4$ & $c_{\Lambda}=1$ & $c_{\Lambda}=1/4$ \\
\hline
$T_{s}/T_{c}$ & -0.89 & -0.89 & -0.84 & -0.61 \\
\hline
$\lambda$ & 19.4 & 18.64 & 11.43 & 3.43 \\
\hline
\end{tabular}\\
\end{center}
where it can be seen that the NLA model with $c_{\Lambda}=4$
is very close to the HTL model. 
The fits to the entropy data (shown in Fig.~\ref{fig:S1}) all lie in
a narrow band for all the models considered. The 
fitted effective coupling $G^2$ for the various models is shown in 
figure \ref{fig:G1}; for comparison, also the 2-loop perturbative 
running coupling in $\overline{\hbox{MS}}$ 
is shown, 
where the renormalization scale is varied between $\pi T$ and $4 \pi T$
and following Ref.~\cite{Gupta:2000hr} we have chosen 
$T_c=0.49 \Lambda_{\overline{\hbox{\scriptsize MS}}}$. 
As can be seen from the plot, the results for the effective 
coupling are well within the range of the 
2-loop perturbative running coupling (for the case $c_{\Lambda}=1/4$
and renormalization scale $\pi T$ the results even seem to be identical,
which is, however, probably only a coincidence).
We also show the result for the coupling obtained
in the semiclassical approach of Ref.~\cite{Schneider:2003uz}\footnote{%
For finite $\mu$ and constant temperatures,
 however, the coupling obtained in \cite{Schneider:2003uz}
rises while our results indicate a decrease of
the coupling (which is consistent with the standard QCD running coupling
with renormalization scale proportional to $\sqrt{T^2+(\mu/\pi)^2}$).}.

In general one can see that the effective coupling becomes bigger when
$c_{\Lambda}$ gets smaller; this is because the hard masses (for equal
values of the coupling) are 
smaller than the soft masses which makes the entropy increase when the
hard parts become more important. Accordingly,
the coupling has to rise in order for the entropy to match the data 
(therefore, for the extreme case $c_{\Lambda}=0$ one finds huge values
of the effective coupling constant).

\begin{figure}
\begin{minipage}[t]{.48\linewidth}
\includegraphics[width=\linewidth]{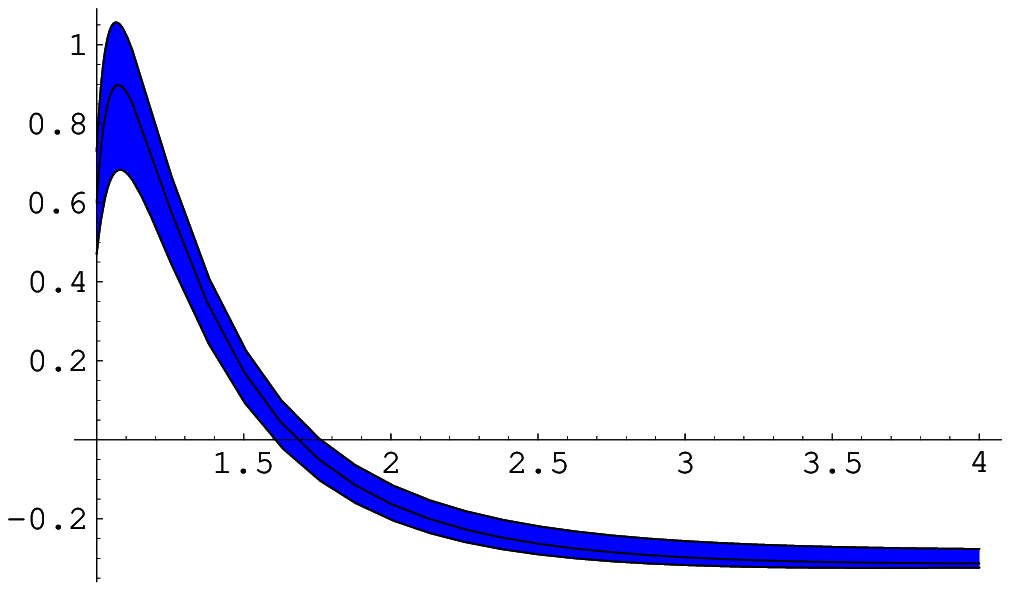}
\setlength{\unitlength}{1cm}
\begin{picture}(5,0)
\put(6.5,2){\makebox(0,0){\footnotesize $T/T_{c}$}}
\put(0,5.7){\makebox(0,0){\footnotesize $B(T,\mu=0)/T^4$}}
\end{picture}
\caption{Residual interaction $B$: NLA model results between
$c_{\Lambda}=4$, $c_{\Lambda}=1$ (full line) 
and $c_{\Lambda}=1/4$ (lower boundary).}
\label{fig:B1}
\end{minipage}\hfill
\begin{minipage}[t]{.48\linewidth}
\includegraphics[width=\linewidth]{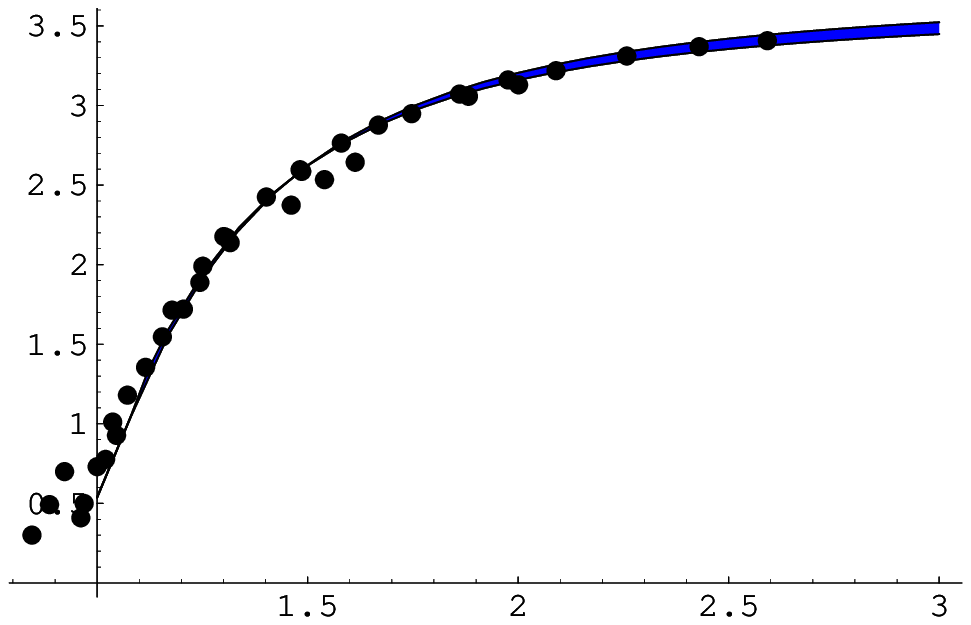}
\setlength{\unitlength}{1cm}
\begin{picture}(6,0)
\put(6.5,0.1){\makebox(0,0){\footnotesize $T/T_{c}$}}
\put(0,5.7){\makebox(0,0){\footnotesize $P/T^4$}}
\end{picture}
\caption{Pressure data from Ref.\cite{AliKhan:2001ek} vs. model pressure.
The band shows the NLA model results for $c_{\Lambda}$ between 
$4$ and $1/4$.}
\label{fig:P1}
\end{minipage}\hfill
\end{figure}

Once the effective coupling is known for $\mu=0$ one 
can proceed to evaluate the QP pressure $\sum p_{i}$ and 
the residual interaction $B$ from Eq.~(\ref{Bdet}), shown in 
Fig.~\ref{fig:B1}; in this figure, the HTL results (not shown) would lie 
marginally above the upper boundary of the band 
(cf. \cite{Romatschke:2002pb} for a comparison between 
simple QP and HTL model).
A comparison between the full pressure (\ref{Pressureansatz}) 
and the lattice data is shown in figure \ref{fig:P1},
where the integration
constant $B_{0}$ was set so that $P(T_{c})=0.536(1)$:\\
\begin{center}
\begin{tabular}[c]{|c|c|c|c|c|}
\hline
& HTL & $c_{\Lambda}=4$ & $c_{\Lambda}=1$ & $c_{\Lambda}=1/4$ \\
\hline
$B_{0}/T_{c}^4$ & 0.82 & 0.73 & 0.6 & 0.47 \\
\hline
\end{tabular}\\
\end{center}

\section{Finite chemical potential}

A calculation and subsequent evaluation
of the coefficients $a_{\mu}$, $a_{T}$ and $b$ in
(\ref{Andrefloweq}) shows that $b$, in contrast to $a_{\mu}$ and $a_{T}$,
is strongly model-dependent. 
Solving the flow equation (\ref{Andrefloweq}) and integrating $B$ along
the characteristics using Eq.~(\ref{Bdet})
one obtains pressure and effective coupling in the whole $T,\mu$ plane.
Here we focus on the pressure but 
the calculation of the other thermodynamic quantities 
is straightforward once 
the effective coupling and the pressure 
have been obtained.

Concerning the shape of the characteristics, it has
already been noticed in \cite{Romatschke:2002pb} that the crossing of 
characteristics
in the simple QP model \cite{Peshier:1999ww}
does not occur in the HTL-model;
this feature is preserved in the NLA models.

\subsection{Lines of constant pressure}

\begin{figure}
\begin{minipage}[t]{.48\linewidth}
\includegraphics[width=\linewidth]{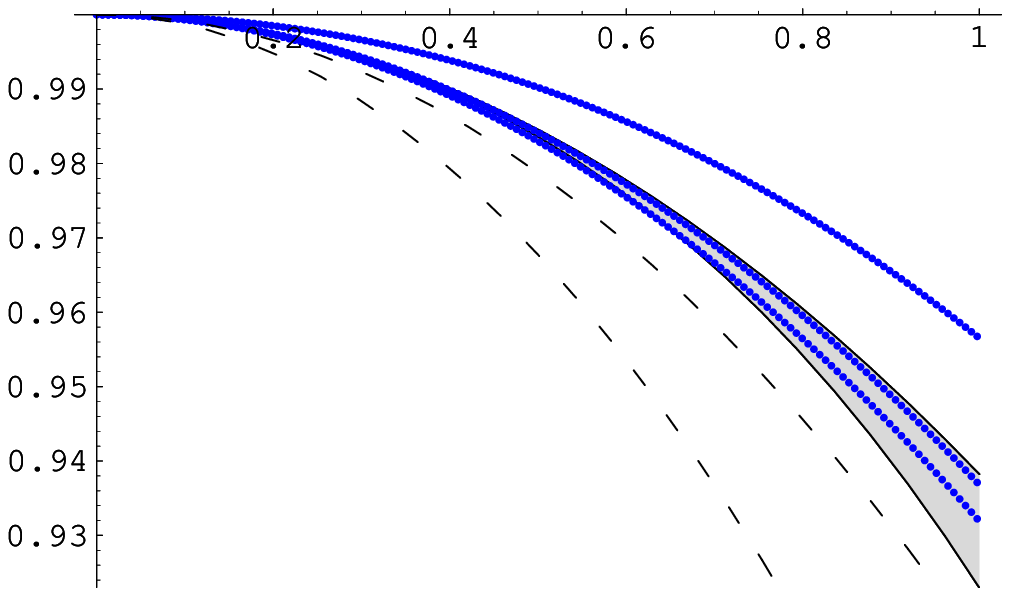}
\setlength{\unitlength}{1cm}
\begin{picture}(6,0)
\put(6.5,5.7){\makebox(0,0){\footnotesize $\mu/T_{c}$}}
\put(0,0){\makebox(0,0){\footnotesize $T/T_{c}$}} 
\end{picture}
\caption{Lines of constant pressure: NLA models for $N_{f}=2$
from $c_{\Lambda}=4$, 
$c_{\Lambda}=1$ and $c_{\Lambda}=1/4$ (dotted lines from lowest to highest);
lattice data for $N_{f}=2+1$ \cite{Fodor:2002km} (light gray band)
and $N_{f}=2$ \cite{Allton:2002zi} (long dashed-lines).}
\label{fig:constP}
\end{minipage} \hfill
\begin{minipage}[t]{.48\linewidth}
\includegraphics[width=\linewidth]{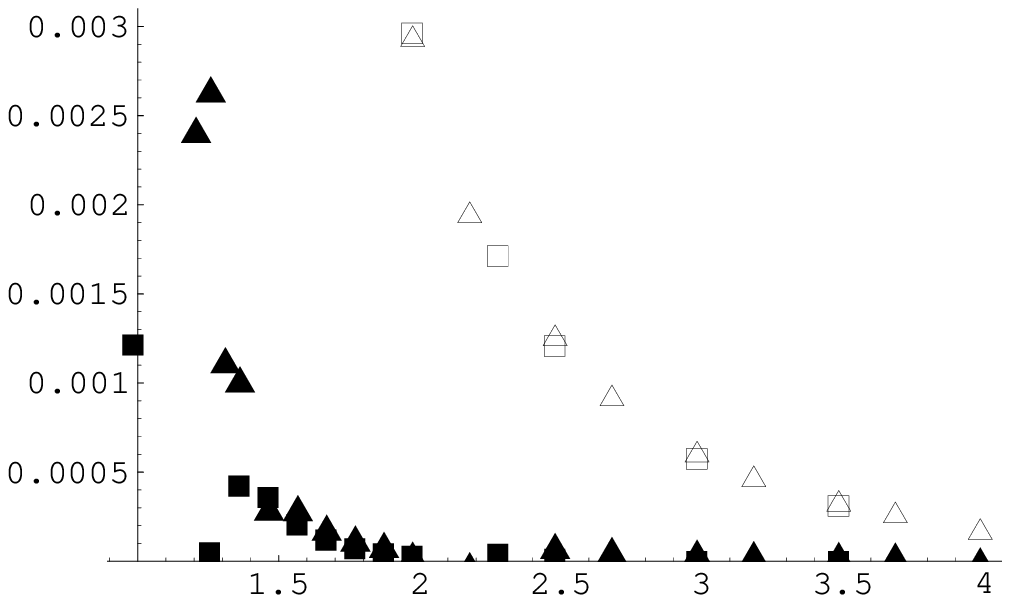}
\setlength{\unitlength}{1cm}
\begin{picture}(6,0)
\put(6.5,0.1){\makebox(0,0){\footnotesize $T/T_c$}}
\put(0,5.7){\makebox(0,0){\footnotesize $\delta P/T^4$}}
\end{picture}
\caption{Difference $\delta P$
scaled with $T^4$ for $\mu/T_{c}=1$. Shown are HTL model (triangles)
and NLA models (boxes). See text for details.}
\label{fig:Chisubst1}
\end{minipage}\hfill
\end{figure}

It is straightforward to extract from our data the line where 
$P(T,\mu)=P(T_{c},0)$; for $N_{f}=2$, the results for 
the simple and HTL QP model as well as for
the NLA models for $c_{\Lambda}>1$ are rather close
to each other. 
For NLA models with $c_{\Lambda}<1$, however, the slope of the 
constant pressure line (until $\mu/T_{c}=1$) deviates rapidly from the
values found for $c_{\Lambda}>1$:
\begin{center}
\begin{tabular}[c]{|c|c|c|c|c|}
\hline
&HTL & $c_{\Lambda}=4$ & $c_{\Lambda}=1$ & $c_{\Lambda}=1/4$ \\
\hline
$T_c \frac{dT}{d\mu^2}$ & -0.06818(8) & -0.06810(6) 
& -0.06329(34) & -0.041(9) \\
\hline
\end{tabular}\\
\end{center}
We have also compared these results with recent lattice data for 2+1 flavors
\cite{Fodor:2002km} in figure \ref{fig:constP}. 
The variations of the lattice 
data in this figure result from fitting the data with a 2nd order
(upper limit) and 4th order polynomial (lower limit) as used in
\cite{Szabo:2003kg}. The comparison shows that our 2-flavor results obtained
from simple, HTL and NLA QP models with $c_{\Lambda}>1$ (indicated
in figure \ref{fig:constP} by the two lower lying pointed lines) 
are in very good
agreement with the lattice data whereas for NLA models with $c_{\Lambda}<1$
the slope is much flatter. Also shown are the lattice results for the
constant pressure line from \cite{Allton:2002zi} for 2 flavors, with 
a slope of $T_c \frac{dT}{d\mu^2}=-0.107(22)$, which is significantly
steeper.

\subsection{Susceptibilities}

It been noticed in \cite{Fodor:2002km} that the quantity
\begin{equation}
\frac{\Delta P}{\Delta P_{SB}}=\frac{P(T,\mu)-P(T,0)}{P_{SB}(T,\mu)-%
P_{SB}(T,0)}
\end{equation}
is essentially $\mu$-independent; we recover a similar scaling behavior
for our models.
As expected, for small chemical potential 
this curve is very close to the quark-number susceptibilities 
$\chi(T)/\chi_0(T)$
obtained at $\mu=0$. In fact, we found that also for larger $\mu$ 
the pressure can be well approximated by  
\begin{equation}\label{Pdiff}
P(T,\mu)-P(T,0)= \frac{\chi(T)}{2} (\mu^2+\frac{\mu^4}{2 \pi^2 T^2}).
\label{dPchieq}
\end{equation}
To quantify this assertion we consider the subtracted quantity
$\delta P=P(T,\mu)-P(T,0)-\frac{\chi(T)}{2}(\mu^2+\frac{\mu^4}{2 \pi^2 T^2})$,
which should be zero, if the pressure at finite chemical potential was 
fully determined by the susceptibilities at $\mu=0$
according to (\ref{Pdiff});
the results (scaled with $T^4$) are shown in 
figure \ref{fig:Chisubst1} for $\mu/T_{c}\simeq 1$. As can be seen,
dropping the $\mu^4$ term in Eq.~(\ref{dPchieq}) (open symbols) deteriorates
the result considerably. Trying to determine the higher-order susceptibility,
$\bar{\chi}=\frac{\partial^4 P}{\partial \mu^4}$ which we parameterized
as
$$
\bar{\chi}=\frac{\chi(T)}{2 T^2} \frac{4!}{x(T)^2},
$$
we find that -- within our numerical errors -- the result for $x(T)$ is
consistent with $\sqrt{2} \pi=4.44\dots$ for $T>1.5 T_{c}$ (which was 
our original assumption in Eq.~(\ref{dPchieq}));
therefore, our results are
close to but about $10$ percent lower than the 
values found in Ref.~\cite{Gavai:2003mf}.

Clearly, $\delta P$ gets smaller with increasing
temperature, but also at relatively low temperatures and high chemical
potential, $\delta P$ is remarkably small (e.g.,
for the HTL model for $T=1.5 T_{c}$ 
and $\mu/T_{c}\simeq 4$ we find a $\delta P/T^4$ smaller than $10$ percent). 

\begin{figure}
\begin{minipage}[t]{.48\linewidth}
\includegraphics[width=\linewidth]{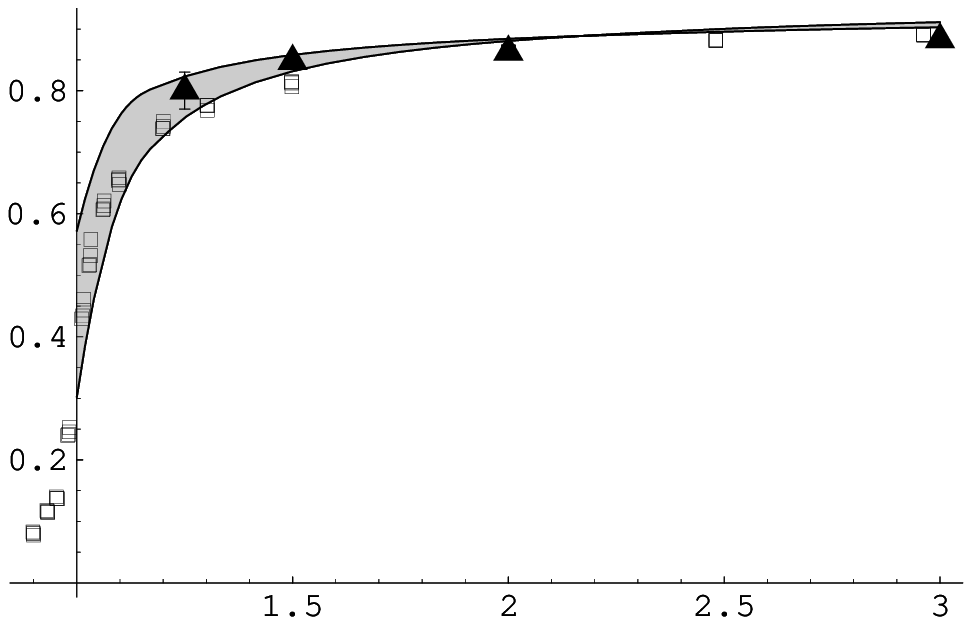}
\setlength{\unitlength}{1cm}
\begin{picture}(6,0)
\put(6.5,0.1){\makebox(0,0){\footnotesize $T/T_{c}$}}
\put(0,5.7){\makebox(0,0){\footnotesize $\chi/\chi_{0}$}}
\end{picture}
\caption{Susceptibilities from NLA models (light gray band), 
from $N_{f}=2$ lattice data \cite{Gavai:2001ie} (triangles) and 
``scaling curve'' for 2+1 flavors from \cite{Fodor:2002km} (boxes).}
\label{fig:Suscep1}
\end{minipage}\hfill
\begin{minipage}[t]{.48\linewidth}
\includegraphics[width=\linewidth]{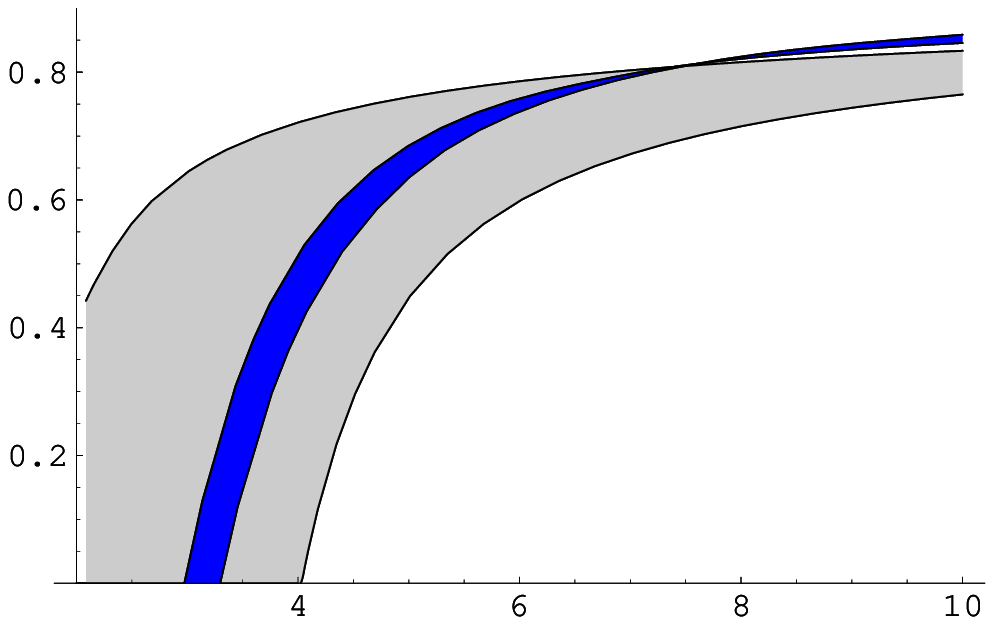}
\setlength{\unitlength}{1cm}
\begin{picture}(6,0)
\put(6.5,0){\makebox(0,0){\footnotesize $\mu/T_{c}$}}
\put(0,5.7){\makebox(0,0){\footnotesize $P/P_{SB}$}} 
\end{picture}
\caption{The pressure at vanishing $T$: Shown are perturbative
results (light gray band) and NLA models for $c_{\Lambda}$ from $1/4$
to $4$ (dark band).}
\label{fig:PT0comp1}
\end{minipage} \hfill
\end{figure}

In Fig.~\ref{fig:Suscep1} we compare the results for the
susceptibilities of our NLA models with the lattice
data for 2 flavors from Ref.~\cite{Gavai:2001ie}
and the ``scaling curve'' in
\cite{Fodor:2002km}, 
which provides an approximation to the 2+1 flavor quark-number
susceptibility at $\mu=0$. We find that all these results
agree very well, with the agreement getting better at
higher temperatures.

It should be noted however that none of the lattice data 
we are using has yet been
rigorously extrapolated to the continuum limit, so that both
the data and the extrapolation by means of QP models are still
likely to change somewhat when this will be done eventually.

\subsection{Very small temperatures}

Extending the lines of constant pressure from our models to very small 
temperatures we obtain a crude estimate of the phase transition line in
this region of phase space. Denoting with $\mu_{c}$ 
the chemical potential where (for vanishing temperature) the pressure 
equals the lattice pressure at $\mu=0$, $P(0,\mu_{c})=P(T_{c},0)$, we
find for our models (assuming $T_{c}=172$ MeV)
\begin{center}
\begin{tabular}[c]{|c|c|c|c|c|}
\hline
& HTL & $c_{\Lambda}=4$ & $c_{\Lambda}=1$ & $c_{\Lambda}=1/4$ \\
\hline
$\mu_{c}$&533 MeV& 536 MeV & 558 MeV & 584 MeV \\
\hline
$\mu_{0}$&509 MeV& 511 MeV & 537 MeV & 567 MeV \\
\hline
\end{tabular}\\
\end{center}
Here we have also given the values $\mu_0$ where the pressure
vanishes, which may be taken as a definite lower bound for the
critical chemical potential within the respective models.

In general, these results are in agreement with the estimates 
for $\mu_{c}$ from \cite{Szabo:2003kg,Peshier:2002ww}; for the QP models
considered, the lowest and highest results for $\mu_{c}$ are obtained
for the HTL model and the NLA model with $c_{\Lambda}=1/4$, respectively,
while the $\mu_{c}$ of the simple QP model lies between the  
NLA $c_{\Lambda}=4$ and $c_{\Lambda}=1$ model values.
However, even our lowest result for $\mu_{c}$ turns out to exceed
the value for the critical chemical potential expected in
Ref.~\cite{Fraga:2001id,Fraga:2001xc}. 

\begin{figure}
\begin{minipage}[t]{.48\linewidth}
\includegraphics[width=\linewidth]{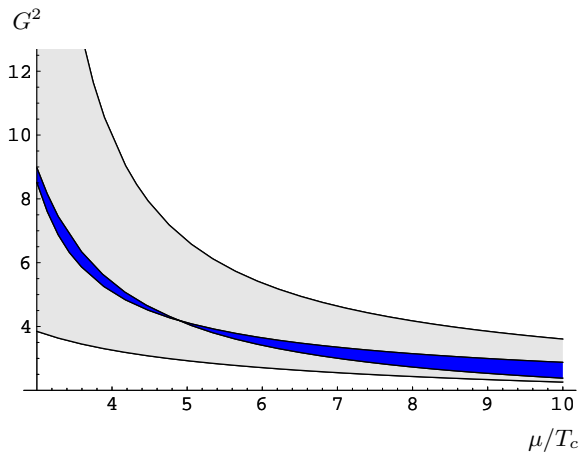}
\setlength{\unitlength}{1cm}
\begin{picture}(6,0)
\put(6.5,0.1){\makebox(0,0){\footnotesize $\mu/T_{c}$}}
\put(-0.5,5.7){\makebox(0,0){\footnotesize $G^2$}}
\end{picture}
\caption{Effective Coupling from NLA models with $c_{\Lambda}=4$ to $1/4$
(dark band) and perturbative 2-loop running coupling (light gray band).} 
\label{fig:G2T0comp}
\end{minipage}\hfill

\end{figure}

A comparison of the perturbative pressure \cite{Freedman:1977ub} 
at vanishing temperature with the results for our models 
at $T/T_{c}\simeq 0.01$ is shown in figure \ref{fig:PT0comp1}. 
We plotted the perturbative results converted to $\overline{\hbox{MS}}$
with the standard 2-loop running coupling, 
$\Lambda_{\overline{\hbox{\scriptsize MS}}}={T_{c}}/{0.49}$ 
\cite{Gupta:2000hr} and renormalization scale varied from $\mu$ to $3\mu$; 
a comparison of this coupling and our results is shown in figure
\ref{fig:G2T0comp}.

\subsection{Equation of state for cold dense matter}

By calculating the number density at $T/T_{c}\simeq 0.01$ and using
our results for the pressure we obtain 
an equation of state for cold deconfined matter. As is the case for 
the simple QP model \cite{Peshier:2002ww}, we find that 
the energy density $\mathcal{E}$ is well fitted
by the linear relation 
$$
\mathcal{E}(p)=4 \tilde{B}+ \alpha p
$$
with
\begin{center}
\begin{tabular}[c]{|c|c|c|c|c|}
\hline
& HTL & $c_{\Lambda}=4$ & $c_{\Lambda}=1$ & $c_{\Lambda}=1/4$ \\
\hline
$4 \tilde{B}/T_{c}^4$ & 11.1(8)&12.3(8)& 14.7(9) & 19.2(1.6) \\
\hline
$\alpha $ &3.23(5)&3.22(4)& 3.22(4)& 3.17(4)\\
\hline
\end{tabular}\\
\end{center}
A value of $\alpha\simeq 3.2$ for $N_{f}=2$ 
has also been found in the simple quasiparticle model
in \cite{Peshier:2002ww} and thus
seems to be model independent, in contrast to the bag 
constant $\tilde{B}^{1/4}$, which varies between $314$ and $360$ MeV.
By using the Tolman-Oppenheimer-Volkov equations and the equation of state
one can determine the mass-radius relations of non-rotating
quark-stars \cite{Andersen:2002jz}; choosing $\alpha=3.2$ and taking the 
bag constant values from above, we find stable star configurations
with radii ranging from $3.6$ to $4.9$ km and masses from $0.7$-$0.8$ solar
masses for NLA, $c_{\Lambda}=1/4$ and HTL models, respectively. 
However, it should be kept in mind that the outermost
layers of such a quark star are metastable with respect to hadronic matter
and therefore the details of the star structure will depend sensitively
on the hadronic equation of state \cite{Peshier:2002ww}.

\section{Conclusions and Outlook}
We considered an improvement of simple quasiparticle models 
\cite{Peshier:1999ww}
by using the full HTL approximation and certain NLO corrections
thereof along the lines of Ref.~\cite{Blaizot:1999ip,Blaizot:2000fc,Blaizot:2003tw} and investigated
the respective predictions at finite chemical potential 
when these models are
matched to $N_{f}=2$ lattice data at $\mu=0$. 

We found that the slope of constant pressure for those models with
$c_{\Lambda}>1$ agrees very well with recent lattice data for 2+1 flavors
\cite{Fodor:2002km,Szabo:2003kg}, while 
the lattice data for $N_f=2$ \cite{Allton:2002zi} 
indicate significantly steeper slopes.

We also found that for our models the pressure for non-vanishing $\mu$
scales with the quark number susceptibilities at $\mu=0$, provided that 
the $\mu^4$ terms (corresponding to a higher order susceptibility) 
are not dropped but added with the same weight as
in the ideal gas limit. We therefore interpreted the 
``scaling curve'' from 2+1 flavor lattice data
\cite{Fodor:2002km} as susceptibilities at $\mu=0$ and found that
a comparison with $N_{f}=2$ susceptibilities from another lattice group 
\cite{Gavai:2001ie} as well as those obtained from our models all agree
very well.

We finally extended our results to the small temperature, high density regime
and obtained an estimate for the critical density at $T=0$ which is consistent
with earlier results, though in excess of the expectations of
Ref.~\cite{Fraga:2001id,Fraga:2001xc}. Furthermore, we obtained an 
equation of state for cold, dense matter, which allows for pure quarks
stars with masses of $\sim 0.8 M_\odot$ and radii of less than $5$ km,
similar to the results obtained by \cite{Fraga:2001xc,Peshier:2002ww,
Blaschke:1998hy}. 


\acknowledgments

We want to thank Z.~Fodor and K.K.~Szabo for 
kindly providing their lattice data
and A.~Peshier, C.~Schmidt and R.A.~Schneider for discussions.
This work has been supported by the Austrian Science Foundation FWF,
project no. 14632.


\end{document}